\documentclass[pra,twocolumn,superscriptaddress,floatfix,amsmath,nofootinbib,amssymb]{revtex4-1}

\usepackage{graphicx}% Include figure files
\usepackage{bm}% bold math
\usepackage{verbatim}
\usepackage{mathrsfs}
\usepackage{color}
\usepackage{paralist}
\usepackage{hyperref}
\usepackage{subfigure}
\usepackage{bbold}
\usepackage{amsthm}

\newcommand{\ii}{\mathrm{i}}

\newcommand{\ket}[1]{| {#1} \rangle}

\newcommand{\bra}[1]{\left\langle {#1} \right|}

\newtheorem{mydef}{Definition}
\newtheorem{thm}{Theorem}

\begin{document}

\title{Work extraction using Gaussian operations in non-interacting fermionic systems}
\author{Marvellous Onuma-Kalu}
\email{monumaka@uwaterloo.ca}
\affiliation{Department of Physics \& Astronomy, University of Waterloo,  Ontario Canada N2L 3G1}
\author{Robert B. Mann}
\email{rbmann@uwaterloo.ca}
\affiliation{Department of Physics \& Astronomy, University of Waterloo,  Ontario Canada N2L 3G1}
\begin{abstract}
We investigate work extraction from  non-interacting fermions  under arbitrary unitary operations and the
more restricted class of  Gaussian unitary operations 
that can be feasibly implemented. 
We characterize general quantum states in fermionic systems according to their ability to yield work (or not) under such  transformations and study  the limit for which multiple copies of passive states in fermionic systems can be activated for work extraction. We find that a sufficient number of copies of  non-thermal passive states
can achieve this, yielding an upper bound on the number of copies needed.
\end{abstract}

\maketitle

\section{Introduction}
 Since the inception of quantum thermodynamics, one of the important areas of research is the search for minimal resources and the least work-intensive protocols for the extraction of work out of thermal systems. This task entails finding quantum states that are freely available and quantum operations that can be feasibly implemented. In this context, the classification of passive states \cite{Pusz1978,Lenard1978} as ``freely available" \cite{fernando2014,carlo2017b}, may have been overrated; this is because over a long period, ``passive states" were generally believed to have no extractable work under  cyclic unitary transformations.  Surprisingly, recent studies   \cite{alicki2013,carlo2017,CPV2015} have shown that this situation holds if we have access to only a single copy of the state. However if we collectively process many copies of the same system, extractable work can become available. If no work can be extracted unitarily, no matter how many copies are available then the state is said to be completely passive; thermal states are the only completely passive states \cite{Pusz1978,Lenard1978}. The fact that passive states can be ``activated'' in such a way that work can be extracted from them is drawing increasing interest amongst researchers \cite{Hov2013,alicki2013,Pera2015,kamil2016}. In this context, it seems that the underlying entanglement structure of the quantum system plays a crucial role. Indeed, recent results call into question the role of entanglement, free energy, correlations and coherence for such work extraction.

Ideally, work can be extracted from non-passive states whose average energy can be lowered by acting on it with  cyclic unitary operations. Generally  in a cyclic Hamiltonian process \cite{CPV2015}, the maximal extractable work (called the ergotropy)  between states $\rho$ and $\sigma$ is given as
\begin{align}\label{maxwork}
W_{\text{max}}(\rho) = \text{max}_{U} \operatorname{tr}\Big[H (\rho - \sigma)\Big]
\end{align}
where $H$ is the system's Hamiltonian, $\sigma = U \rho U^{\dagger}$, and $U$ the unitary operator. For $n$ copies of passive states, $U$ would be a global (entangling) unitary acting on the total system with total Hamiltonian given by $H$. 

 Global (and thus entangling) unitary operations are capable of extracting more work than local operations from a set of quantum systems. However, the dynamics involving a global operation is slow, in the sense that it requires many different operations. Since such operations are difficult to implement, we are therefore left to consider which work extraction protocol is practically achievable when subjected to a restricted unitary operation. A large class of transformations that are easy to describe are Gaussian unitaries, which map Gaussian states into Gaussian states. The Gaussian unitary transformations being generated by quadratic Hamiltonians are in general more constraint than general unitary transformations. 

A characterization of bosonic quantum states from which no (or maximal) work can be extracted using a Gaussian unitary transformation was recently established \cite{eric2016}. In this context, bosonic Gaussian passive states (and non-Gaussian passive states), from which no (or maximal) work can be extracted using a Gaussian unitary transformation, were defined. In this paper we investigate the corresponding situation for fermionic systems. 
Fermionic systems are similar to bosonic systems but differ in their statistics (Fermi-Dirac in the former and Bose-Einstein in the latter). There is a one-to-one map between $n$ fermionic modes and the Hilbert space of $n$ qubits. This allows for easy computations with fermions, providing  an added advantage for  quantum computational tasks \cite{Bravyi2000}.  It is the main aim of this paper to see how useful a fermionic system is for work extraction.  We will show how Gaussian unitaries can yield a characterization of Gaussian passive and Gaussian non-passive fermionic states respectively.

Energy storage and its subsequent extraction has both fundamental and practical importance.  The main goal of our study of work extraction from non-interacting fermion systems is to understand from which quantum states of fermionic systems energy can or cannot be minimized. We first consider general unitary transformations  and 
then investigate the more restricted class of Gaussian unitary operations. Specifically we consider a $1D$
non-interacting continuous variable fermionic systems composed of $n$ modes. 

In section \ref{tu}, we describe the features of such systems and then discuss the characterization of fermionic Gaussian states. Identical thermal states do not allow for work activation no matter how many copies are available since a product of thermal states is a thermal state itself and hence passive. As a proof, we give an illustration of activation in product of thermal states using a general unitary transformation  in section \ref{twoo} . Our main results are presented in section \ref{four} where we characterize Gaussian and non-Gaussian passive states based on the availability of extractable work using Gaussian operations.

\section{Continuous Variable Fermion Systems}\label{tu}

A continuous variable (CV) system of $N$ canonical fermionic modes $\kappa$ is described by a Hilbert space (with dimension $2^{N}$) $\mathcal{H} = \bigotimes_{\kappa = 1}^{N}\mathcal{H}_{\kappa}$ spanned by  the basis $\ket{n_{\kappa}=0}$ and  $ \ket{n_{\kappa}=1}$ known as the Fock or number state basis. The annihilation and creation operators $a_{k}$ and $a_{k}^{\dagger}$ of a fermionic particle in mode $k$ (with frequency $\omega_{k}$) satisfy the canonical anti-commutation relation (CAR) $\lbrace a_{k}^{\dagger},a_{l}\rbrace = \delta_{kl}, \quad \lbrace a_{k}^{\dagger},a_{l}^{\dagger}\rbrace = \lbrace a_{k},a_{l}\rbrace =0 $, where $\delta$ is the Kronecker delta.   Over the Fock state, the action of $\hat{a}$ and $\hat{a}^{\dagger}$ operators is given as $a\ket{0} = 0 = a^{\dagger}\ket{1}$, $a\ket{1} = \ket{0}$, and  $a^{\dagger}\ket{0}=\ket{1} $. The number operator $\hat{n} = a^{\dagger}a$ is an eigenstate of the Fock state, i.e $n\ket{n} = n\ket{n}$.
 
The fermionic system may be described by another set of field operators known as the Majorana operators. The $kth$ mode Majorana operators $c_{2k}$ and $c_{2k-1}$ are related to the creation and annihilation operators through the relation 
\begin{align}\label{maj}
c_{2k - 1} = \frac{1}{\sqrt{2}} (a_{k} + a^{\dagger}_{k}), \qquad c_{2k} = \frac{\ii}{\sqrt{2}}(a_{k} - a_{k}^{\dagger}),
\end{align}
where
$k=1,2,\cdots ,N$ labels the $N$ modes of the system under study. The Majorana operators are Hermitian and satisfy the relation $\lbrace c_{i},c_{j} \rbrace = \delta_{ij}$. They can be arranged into a vector   
\begin{align*}
\mathbf{\hat{x}}:= (c_{1},c_{3},\cdots, c_{2M-1}; c_{2},c_{4},\cdots, c_{2M})^{T}
\end{align*}
so that in a compact form, the fermionic canonical anticommutation relations (CAR) become
\begin{align}
\lbrace \hat{x}_{i} , \hat{x}_{j}\rbrace = \delta_{ij}.
\end{align}

Linear transformations on fermionic operators that preserve the CAR are of the form
\begin{align*}
c_{k} \rightarrow c'_{k} = \sum_{i}O_{kl}c_{l}
\end{align*}
where $O \in O(2M)$ is an element of the orthogonal group. These transformations can be implemented by unitary operations which are generated by quadratic Hamiltonians in the Majorana operators $c_{j}$.  

%We can also define a set of anticommuting variables $\gamma_{i}$ and their conjugates $\gamma_{i}^{*}$, which obey the relations
%\begin{align*}
%\lbrace \gamma_{i}^{*},\gamma_{j}\rbrace = \lbrace \gamma_{i}^{*},\gamma_{j}^{*}\rbrace= \lbrace \gamma_{i},\gamma_{j}\rbrace = 0
%\end{align*}
%and anticommutes with fermionic operators $\lbrace \gamma_{i},a_{j}\rbrace = 0$. The independent variables $\gamma_{i}$ and $\gamma_{i}^{*}$ are called Grassmann variables \cite{Cahill1999}.
%

\subsection{Fermionic Gaussian States}\label{FGx}
Gaussian states are easily accessible in laboratories and Gaussian unitaries can be easily implemented. The idea that the unitary operation necessary to extract work from passive states is rather general led to considerations of the set of more restricted set of Gaussian transformations as they are more practically implemented \cite{eric2016}.  This brought the notion of bosonic Gaussian passive states as those states from which work cannot be extracted using Gaussian transformation in the bosonic regime \cite{eric2016}.

In the sequel we shall consider the analogous problem for fermions.  To this end, we recapitulate some basic
formalism on fermionic Gaussian states, which may be defined based on either covariance matrix approach or a Grassmann approach \cite{Cahill1999}.

\textit{Covariance matrix approach:} Here, arbitrary fermionic Gaussian states are operators that are exponentials of quadratic form in the Majorana operators
\begin{align}\label{FGS}
\rho = Z^{-1}  \exp\Big[\frac{-\ii}{4}\mathbf{x}^{T}G\mathbf{x} \Big],
\end{align}
where $Z$ is a normalization constant. $G$ is  a real anti-symmetric $2M \times 2M$ matrix, which can be brought to a $2M\times 2M$ block diagonal form by a special orthogonal matrix $O \in SO(2M)$, that is 
\begin{align}\label{trans}
\tilde{G}= O G O^{T} = \bigoplus_{j=1}^{M} \begin{pmatrix} 0& \beta_{j}\\-\beta_{j}& 0 \end{pmatrix}
\end{align}
where the $\beta_{j}$ are real numbers that characterize $G$. With an inverse transformation $G=O^{T}\tilde{G}O$, the density matrix \eqref{FGS} can be written as
\begin{align}\label{FGS1}
\rho = Z^{-1} \exp\Big[-\frac{\ii}{4}\mathbf{x}^{T}O^{T}\tilde{G}O\mathbf{x} \Big] 
\end{align}
and upon defining a new set of transformed Majorana operators $\tilde{\mathbf{x} }= \mathbf{O} \mathbf{x}$, the density matrix \eqref{FGS1} becomes
\begin{align*}
\rho &= Z^{-1}\exp\Big[-\frac{\ii}{4}\tilde{\mathbf{x}}^{T}\tilde{G}\tilde{\mathbf{x} }\Big]
\end{align*}
Substituting $\tilde{\mathbf{x} } = (\tilde{c}_{2j-1},\tilde{c}_{2j})^{T}$  and equation \eqref{trans}, we get the fermionic Gaussian state in standard form \cite{B2005}
\begin{align}\label{FGS2}
\rho =   \frac{1}{2^{n}} \Pi_{j=1}^{n}(\mathbb{1} - \ii \tanh\Big( \frac{\beta_{j}}{2}\Big)
 \tilde{c}_{2j-1}\tilde{c}_{2j})
\end{align} 
using $(\tilde{c}_{2j-1}\tilde{c}_{2j})^2 = -1$ for any $j$.  

%where the $\beta_j$ are real numbers \tcr{called the Williamson eigenvalues \cite{} for matrix $G$}.  So through a unitary evolution with Hamiltonians quadratic in the Majorana operators $c_{j}$,  the state \eqref{FGS} in the standard form becomes
%\begin{align}\label{statesf}
%\tcr{\rho_{st}= \frac{1}{2^{n}} \Pi_{j=1}^{n}(\mathbb{1} - \ii \lambda_{j} \tilde{c}_{2j-1}\tilde{c}_{2j})}
%\end{align}
%\tcr{where $\tilde{c} = O^{T}c$ and $\lambda_{j} \in [-1,1]$ are real numbers. We can think of the state in standard form as having the unitary operation decouple the state modes and transforming the state into one that is a product of independent thermal states diagonal in the number basis.} 

 Let us define a  real and anti-symmetric matrix $\Gamma$ with elements
\begin{align}\label{cm}
\Gamma_{kl} =  \frac{\ii}{2}\langle [c_{k}, c_{l}]\rangle  = \begin{cases} \ii  \langle c_{k}c_{l}\rangle, & \text{for}\ k \neq l\\ 0, & \text{for} \ k = l \end{cases}
\end{align}
where for a given state $\rho$ and an observable $A$, we define $\langle A \rangle = \operatorname{Tr}[\rho A]$.  In terms of the transformed Majorana operators $\tilde{c} = O c$, the anti-symmetric matrix $\Gamma$ transforms as
\begin{align*}
\tilde{\Gamma}_{kl} &= \frac{\ii}{2} \operatorname{Tr}(\rho[\tilde{c}_{k}, \tilde{c}_{l}]) = \frac{\ii}{2}\langle [O_{km}c_{m}, O_{ln}c_{n}]\rangle \\
&= \sum_{kl}O_{km}\frac{\ii}{2}\operatorname{Tr}(\rho[c_{m},c_{n}])O_{nl}^{T} = \mathbf{O}\mathbf{\Gamma}\mathbf{O}^{T}
\end{align*}
Using the density matrix from \eqref{FGS2}, we can calculate  $\tilde{\Gamma}_{kl} = \frac{\ii}{2} \operatorname{Tr}(\rho[\tilde{c}_{k},\tilde{c}_{l}])$, obtaining
\begin{align*}
\lambda_{j} = \tilde{\Gamma}_{2j-1,2j} =\ii \operatorname{Tr}(\rho [\tilde{c}_{2j-1},\tilde{c}_{2j}]) = \tanh\Big(\frac{\beta_{j}}{2}\Big)
 \end{align*}
with all other $\tilde{\Gamma}_{kl}$ zero, yielding
\begin{align}
\tilde{\Gamma} = O \Gamma O^{T} = \bigoplus_{j = 1}^{M} \begin{pmatrix} 0& \lambda_{j}\\-\lambda_{j}& 0 \end{pmatrix}
\end{align}
and demonstrating that $\Gamma$ and $G$ can both be brought to block diagonal form by the same orthogonal matrix $\mathbf{O}$.   

By definition, $\Gamma$ is the covariance matrix of the fermionic Gaussian state. The direct link between $G$ and $\Gamma$ indicates that a fermionic Gaussian state can be fully characterized by either its density matrix or its covariance matrix. Hence every $\Gamma$ corresponding to a physical state has to fulfil $\ii  \Gamma \leq \mathbb{1}$ or equivalently $\Gamma \Gamma^{T} \leq \mathbb{1}$ with equality if and only if the state is pure.  

 Thermal states with inverse temperature ($\beta$)
\begin{align*}
\tau (\beta)= \frac{1}{\mathcal{Z}}e^{-\beta H}, 
\end{align*}
are examples of a more general class of Gaussian state. Here $\mathcal{Z}$ is the partition function. If we define the Hamiltonian for a single fermionic mode with frequency $\omega$ as $H=\omega a^{\dagger}a$, then in the Fock basis, a thermal state can be expressed as 
\begin{align*}
\tau(\beta) = \frac{1}{(1+e^{-\beta \omega})} \sum_{n=0}^1e^{-n \beta \omega}\ket{n}\langle n|
\end{align*}
with covariance matrix
\begin{align*}
\Gamma =\begin{pmatrix} 0 & \lambda\\
-\lambda & 0\end{pmatrix}, \quad \Gamma ^{2} < -\lambda^{2}\mathbb{1}
\end{align*} 
where $\lambda = \tanh\Big( \frac{\beta \omega}{2}\Big)$. For $n$ non interacting fermionic modes, the Hamiltonian is defined as $H = \sum_{i=1}^{n} \omega_{i}a^{\dagger}_{i}a_{i}$ and the covariance matrix for the  product of $n$ fermionic thermal states is
\begin{align}\label{therm}
\Gamma_{n}= \bigoplus _{i}^{n} \begin{pmatrix} 0 & \lambda_{i}\\
-\lambda_{i} & 0\end{pmatrix},
\end{align} 
$\lambda_{i} = \tanh\Big( \frac{\beta_{i} \omega_{i}}{2}\Big)$. We will make reference of this later in the paper. 

\textit{Grassmann approach:} The connection between the covariance matrix approach and Grassmann approach is the map assigning the Grassmann variables to each Majorana operator
\begin{align}
\omega(c_{2M-1},c_{2M},\gamma)= \gamma_{2M-1}\gamma_{2M}, \quad \omega(\mathbb{1},\gamma)=1
\end{align}
where $\gamma_{k}\in \mathcal{G}_{2n}$ is the algebra of Grassmann variables. Then we define a state $\rho$ of $n$ fermionic modes to be Gaussian if its Grassmann representation $ \omega(\rho,\gamma)$ is Gaussian
\begin{align}
\omega(\rho,\gamma) = \frac{1}{2^{n}}\exp\Big(\frac{\ii}{2}\gamma^{*}\Gamma \gamma\Big)
\end{align}
where $\Gamma$ is a $2n \times 2n$ real antisymmetric matrix also known as the covariance matrix of the state as defined in \eqref{cm} \cite{B2005}. 

\textit{Coherent states}: Under the Grassmann representation, one can define a fermionic coherent state \cite{Cahill1999}. For any set of variables $\lbrace \gamma_{i}\rbrace$ of Grassmann numbers, a normalized coherent state $\ket{\gamma}$ is defined as the displaced vacuum state $\ket{\gamma }= D(\gamma)\ket{0}$, where $D(\gamma)$ is the displacement operator which acts on fermionic $\hat{a}$ and $\hat{a}^{\dagger}$ operators as $D(\gamma)\hat{a}D^{\dagger}(\gamma) = a + \gamma$ and $D(\gamma)\hat{a}^{\dagger}D^{\dagger} (\gamma)= a^{\dagger} + \gamma^{*}$ respectively \cite{Cahill1999}.This operation preserves the anticommutation relations.  In this paper  we restrict ourselves to  Gaussian operators that contain no  Grassmann variables \cite{corney2006}.

\subsection{ Gaussian unitaries}

 Gaussian unitaries are generated by Hamiltonians quadratic in Majorana operators and transform Gaussian states to Gaussian states. This definition applies to the specific case of unitary transformations that preserve the Gaussian character of a quantum state. Gaussian transformations in Hilbert space are special orthogonal transformations on phase space.  In terms of the statistical moment $\mathbf{\hat{x}}$  and $\Gamma$, the special orthogonal transformation is defined by the action \begin{align}
\mathbf{\hat{x}} = O\mathbf{\hat{x}}, \quad \Gamma = O \Gamma O^{T}, \quad OO^{T} = \mathbb{1}
\end{align}

Unlike boson field operators, whose algebraic properties are preserved by  symplectic transformations, fermion anticommutation relations are invariant under rotations. Examples of such Gaussian transformations  that  preserve the canonical anticommutation relations of fermionic modes (thus transforming fermionic Gaussian states to fermionic Gaussian states) are
as follows. 
\begin{itemize}
\item Phase rotation operator:
\begin{align}
&R(\theta) = e^{-\ii \theta a^{\dagger}a} \nonumber\\
&\mathbf{\hat{x}} \rightarrow O(\theta)\mathbf{\hat{x}} , \quad O (\theta)= \begin{pmatrix}\cos(\theta) & \sin(\theta) \\ - \sin(\theta)& \cos(\theta) \end{pmatrix} 
\label{rot1}
\end{align}
\item Two-mode squeezing operator \cite{pra2007}  
\begin{align}
&S(r) = \exp\Big[ r(ab - b^{\dagger}a^{\dagger} )\Big]\nonumber\\
&\mathbf{\hat{x}} \rightarrow S(r) \mathbf{\hat{x}}, \quad S(r) =  \begin{pmatrix} \cos (r)\mathbb{1} & -\sin (r) \sigma_{z}\\ \sin(r) \sigma_{z} & \cos(r) \mathbb{1}\end{pmatrix}
\label{sq1}
\end{align}
where 
 $\sigma_{z} = \operatorname{diag}(1,-1)$ is the usual Pauli matrix.
\item Beam splitting operation      
\begin{align}
&B(\phi)= \exp\Big[ \phi (ab^{\dagger} + a^{\dagger}b)\Big]
\nonumber\\
& \mathbf{\hat{x}} \rightarrow B(\phi)\mathbf{\hat{x}},\quad B(\phi) = \begin{pmatrix}\cos(\phi) \mathbb{1} & -\sin(\phi) \mathbb{1}\\\\
\sin(\phi) \mathbb{1} & \cos(\phi) \mathbb{1} \end{pmatrix}
\label{bs1}
\end{align}
\end{itemize}

\section{Passivity and Activation}\label{twoo}

 A state $\rho$ is passive if its average energy cannot be lowered when a unitary operation acts on it, that is 
\begin{align}
\operatorname{Tr}[H\rho]\leq \operatorname{Tr}[H U \rho U^{\dagger}],
\end{align}
where $H=\sum_{i=0}^{d-1} E_{i}\ket{i}\langle i|$ is the Hamiltonian of the finite dimensional quantum system associated with the Hilbert space $\mathcal{H}\equiv \mathbb{C}^{d}$, with energy eigenstates $\ket{i}$ and eigenvalues $E_{i}$. A state may be passive given only a single copy but  can become active for $n$ copies. Completely passive states remain passive no matter how many copies of the system are available, while those states that become active for some $k\geq n$ copies of the system is termed $k$-activable \cite{CPV2015}. This  naturally leads to the question of what  the class of states is that remains passive, even given an infinite number of copies. Thermal states defined by $\rho=\frac{1}{\mathcal{Z}}e^{-\beta H}$ with $\mathcal{Z} = \operatorname{Tr}[e^{-\beta H}]$ are the only completely passive states \cite{Pusz1978,Lenard1978}.

Given that some passive states can be activated for some $k\geq n$ copies of the system to yield work, the aim of this section is to find the value of $k$  for which a passive but not thermal state of fermionic modes can be activated to yield work.

\subsection{Passive states}

 Passivity of a quantum state is often expressed as a property of the state and its Hamiltonian. Consider a state $\rho$ and a reference Hamiltonian $H$, both written in their respective eigenbasis,
\begin{align*}
H:= \sum E_{n}\ket{n}\langle n |, \quad \text{with}\quad  E_{n+1} \geq E_{n}\; \forall n,\\
\rho: = \sum p _{n} \ket{\rho_{n}}\langle \rho_{n}|, \quad \text{with} \quad p_{n+1} \leq p_{n}\; \forall n
\end{align*}
where $0\leq p_{n} \leq 1$ and $\sum_{n}p_{n}=1$.  $\rho$ is passive if and only if it is diagonal in the same basis as the Hamiltonian $H$ of the system, that is $[\rho,H]=0$. This can be interpreted as $\lbrace \ket{\rho_{n}} \rbrace$ coinciding with $\lbrace \ket{n}\rbrace$, with no population inversion, that is with decreasing population $p_{j} < p_{k}$ and increasing energy $E_{j} > E_{k}$.  Otherwise we say $\rho$ is non-passive. 

%\tcr{Some states that are passive for a single copy of the state may not be passive for multiple copies of the state \cite{CPV2015} hence exhibiting a form of activation.That is, there exist situations where a unitary $U$ acting on $n$ copies of a system is able to lower the average energy of the total system, whilst if one had access to many copies of the system no such unitary exists. This then naturally leads to the question of what is the class of states which remain passive under composition, i.e from which no work can be extracted even from an infinite number of copies. States that cannot be activated regardless of the number of copies are called completely passive states. } 

 In a two-dimensional continuous variable system spanned by the states $\ket{m}$ and $\ket{n}$, it can be shown\cite{eric2016} that a product of two thermal states of two bosonic modes at the same inverse temperature $\beta$ and frequency $\omega$, form an example of a passive state, whereas given the modes with the same frequency and at different inverse temperature, the state is  non passive. We ask if this is true for fermionic systems.

\subsection{Activation of passive states to generate work}\label{act}

In the Fock basis, a thermal state for a fermionic mode with inverse temperature $\beta$ is given as
\begin{align}\label{taupass}\nonumber
\tau(\beta) &= (1+e^{-\beta \omega})^{-1}\sum_{n=0}^{1}e^{-n \beta \omega}\ket{n}\langle n|\\
&= (1+e^{-\beta \omega})^{-1} (\ket{0}\langle 0| + e^{- \beta \omega}\ket{1}\langle 1|)\
\end{align}

Consider a non-interacting two-mode fermionic system of equal frequency $\omega$ each with local Hamiltonian $h_{i}= \omega a^{\dagger}_{i}a_{i}$. The total Hamiltonian $H$ of the system is simply the sum of the individual local Hamiltonians: $H_{s} = \omega (a_{1}^{\dagger}a_{1} + a_{2}^{\dagger}a_{2})$. The fermionic two-mode thermal state in the Fock basis may then be expressed as
\begin{align*}
\tau(\beta_{1},\beta_{2}) =&\frac{1}{\mathcal{Z}_{1}\mathcal{Z}_{2}}  \sum_{m,n=0}^{1} e^{-\omega (n\beta_{1} + m\beta_{2})}\ket{m}\langle m| \otimes \ket{n}\langle n|
\end{align*}
where $ \mathcal{Z}_{1}\mathcal{Z}_{2} = (1+e^{-\beta_{1} \omega})(1+e^{-\beta_{2} \omega})$ and up to a common factor, the matrix elements are
\begin{align*}
\epsilon = e^{-\omega(\beta_{1}n + \beta_{2}m)} = e^{-\frac{\omega}{T_{1}T_{2}} (mT_{1} + nT_{2})}.
\end{align*}
We see that $H_{s}$ commutes with the product state $\tau(\beta_{1},\beta_{2})$
composed of states of the form in \eqref{taupass}. The occupational numbers $n,m \in \lbrace 0,1 \rbrace$. The sum of the occupational numbers in the state is $N_{i} = m+n$.  Consider a unitary transformation from the state $\tau(\beta_{1},\beta_{2})$ to $\tau'(\beta_{1},\beta_{2})$ such that
\begin{align*}
\tau'(\beta_{1},\beta_{2}) =&\frac{1}{\mathcal{Z}_{1}\mathcal{Z}_{2}}  \sum_{m',n'=0}^{1} e^{-\omega (n'\beta_{1} + m'\beta_{2})}\ket{m'}\langle m'| \otimes \ket{n'}\langle n'|
\end{align*}
with new occupational number given as $N'_{i} = m'+n'$  and matrix element proportional to
\begin{align*}
\epsilon' = e^{-\omega(\beta_{1}n' + \beta_{2}m')} = e^{-\frac{\omega}{T_{1}T_{2}} (m'T_{1} + n'T_{2})}.
\end{align*}
The state $\tau(\beta_{1},\beta_{2})$ is non passive if there exist pairs of non-negative integers $m,n,m',n'$ such that
\begin{align}
\epsilon'>\epsilon, \quad \text{while} \quad  m'+n' > m+ n
\end{align}
which up to a common factor yields the condition
\begin{align}\label{two}
mT_{1} + nT_{2} > m'T_{1} + n'T_{2}, \quad \text{while} \quad  m'+n' > m+ n
\end{align}
by making use of the fact that $e^{-AX} > e^{-AY} \Rightarrow X< Y$.
Given that $m,n \in \lbrace 0,1\rbrace$, equation \eqref{two} cannot be satisfied. We then conclude that for  two-mode fermionic states, regardless of frequencies of the modes and its temperature, the product of two thermal states is always passive, this is in contrast to the bosonic case \cite{eric2016}.

However, for a product $\tau(\beta_{1},\beta_{2},\beta_{3})$  of three fermionic thermal states
\begin{align}\label{state}\nonumber
\tau(\beta_{1},\beta_{2},\beta_{3}) = &  \frac{1}{Z_{1}Z_{2}Z_{3}}  \sum_{m,n,l=0}^{1} e^{-\omega (n\beta_{1} + m\beta_{2}+l\beta_{3})}\\ &\times\ket{m}\langle m| \otimes \ket{n}\langle n| \otimes \ket{l}\langle l|
\end{align}
the situation changes. The non-passivity condition becomes
\begin{align}\label{npfermi}
n\beta_{1}  + m \beta_{2}  + l \beta_{3} &  > n' \beta'_{1}+ m' \beta'_{2} + l' \beta'_{3}\\
 \quad \text{while}\quad  m'+n' +l' & > m+ n+ l \nonumber
\end{align}
The matrix element is now proportional to $e^{-\omega (n\beta_{1} + m\beta_{2}+l\beta_{3})}$ and $m,n,l \in \lbrace 0,1\rbrace$. One can now find a three-dimensional subspace in which a unitary can reduce the average energy, proving that the state $\tau(\beta_{1},\beta_{2},\beta_{3})$ is not always passive. For example, let $m' = n' = 1, l'=0$ and $m=n=0,l=1$, it is obvious that 
 $m'+n'+l' > m+n+l$. Also, 
\begin{align}\label{3-Tcond}
\beta_{3} >  \beta_{1} + \beta_{2}
\end{align}
 which can hold for sufficiently large $\beta_3$. In general the condition \eqref{npfermi} can be satisfied provided   $\beta_{i} \ll \beta_{j},\beta_{k}$ for distinct $i,j,k$. Hence a product of thermal states $\rho = \prod_j^n  \tau(\beta_j) = \tau(\beta_1)\otimes \cdots \otimes \tau(\beta_n) $ for fermionic modes can be activated to become non-passive for $n \geq 3$.  In other words, the state  is $3$-activable \cite{carlo2017}.

From the above we can construct the following \\
\noindent \textit{Protocol:}
Consider the three-mode fermionic system described by the state \eqref{state}. From the non-passivity condition \eqref{npfermi}, we note that for the above transformation to be possible, the action of the unitary operation must be such that
\begin{itemize}
\item The initial state with a composition of the three modes should consists at least of an unpopulated mode and a populated mode. That is, initial states of the system of the form $\ket{111}$ and $\ket{000}$ are not allowed.
\item The action of the unitary should take the initially populated (unpopulated) mode to an unpopulated (populated) mode of the final state.
\item One can always guess the temperature relationship of the different modes: The sum of the inverse temperature of the initially unpopulated modes must be less than the inverse temperature of the populated mode.
\item If a transformation leaves a mode unaffected, then the temperature of such mode does not matter during the transformation process.
\end{itemize}

We now turn to a practical example of such transformation. The three mode state can be written as 
\begin{align*}
\rho_{nml} =& \frac{1}{Z_{1}Z_{2}Z_{3}}  \Big[e^{-\omega \beta_{1}}\ket{100}\langle 001 | +e^{\omega \beta_{3}}\ket{001}\langle 100 |\\
&+e^{-\omega\beta_{2}} \ket{010}\langle 010 | + e^{-\omega(\beta_{1}+\beta_{3})}\ket{101}\langle 101| \\
&+ e^{-\omega(\beta_{1}+\beta_{2})} \ket{110}\langle 011 |  +e^{-\omega(\beta_{2}+\beta_{3})}\ket{011}\langle 110 | \\
&+ \ket{000}\bra{000} + e^{-\omega(\beta_{1}+\beta_{2}+\beta_{3})}\ket{111}\bra{111}\Big]
\end{align*}
upon expanding the sum in \eqref{state}. 
Consider a unitary of the form
 \begin{align}\label{unit}
 U =   \ket{101}\langle 010 | + \ket{010}\langle 101| - \ket{101}\langle 101 | -\ket{010}\langle 010 | +\openone 
 \end{align}
 where $U$ induces a transition between the two degenerate states 
 \begin{align}
 \ket{010} \leftrightarrow \ket{101}
 \end{align}
 We note that $U= U^{\dagger}$. The amount of work extracted from the system (the change in its average energy) is given by \cite{CPV2015}
 \begin{align*}
 W &= \operatorname{Tr}[H(\rho_{nml}-U\rho_{nml} U^{\dagger})] \\
 &= \hbar \omega e^{-\omega \beta_{2}} \Big( 1-e^{-\omega((\beta_{1}+\beta_{3})-\beta_{2})}\Big)
  \end{align*}
which must be positive for the state to be non-passive. Clearly this will hold whenever $(\beta_{1}+\beta_{3})-\beta_{2} <0$ or in other words
  \begin{align}
  \beta_{2} > \beta_{1}+\beta_{3}
  \end{align}
  which agrees with  the non-passivity condition in \eqref{npfermi}. 
 Alternatively one could employ a unitary that interchanges the $    \ket{001}$ and $\ket{110}$ states
 and one would obtain \eqref{3-Tcond}.

  The problem of generating a unitary analogous to \eqref{unit} for more copies of fermion states 
is rather challenging. In the next section, we discuss a more restricted class of unitary transformations.

\section{Gaussian passivity}\label{four}

In the previous section we saw that, unlike the situation for bosonic modes, for fermionic modes a product of two thermal states at different temperatures is passive. Given that  constructing a heat engine requires access to two thermal baths at different temperatures, can one construct a heat engine out of a product of thermal states in fermionic modes? 

To answer this question, we note that passivity of quantum states requires a cyclic unitary transformation. In our work, we consider a Gaussian unitary transformation to characterize fermionic states according to their abilities to generate work or not.

Suppose we have access to a Gaussian unitary. We are  interested in the effect of the Gaussian transformation induced by this unitary on an arbitrary state via the effect on the corresponding covariance matrix. We ask for which (not necessarily Gaussian) states of two non-interacting fermionic modes with frequencies $\omega_{a}$ and $\omega_{b}$ ($\omega_{a} \leq  \omega_{b})$ can energy can be extracted using only Gaussian operations. States from which energy cannot be extracted using Gaussian operations are called Gaussian passive \cite{eric2016}. 

\section{Energy as a function of state covariance matrix}

Before we proceed, we define the average energy of a state in terms of its covariance matrix.
\begin{mydef}
The average energy of a quantum state $\rho$ of a fermionic mode with frequency $\omega$ is given in terms of its covariance matrix $\Gamma$ by the relation
\begin{align}\label{endef}
E(\Gamma) =  \frac{\omega}{2} \Big(1-\operatorname{Tr}(\Omega \Gamma)\Big)
\end{align}
for some real symplectic matrix $\Omega$
\end{mydef}
 To demonstrate this, the covariance matrix $\Gamma$ of a quantum state $\rho$ for a fermionic mode with frequency $\omega$ is
\begin{align}\label{cov}
\Gamma =   \frac{\ii}{2} \begin{pmatrix} 0 &\langle \big[ c_{1},c_{2} \big] \rangle \\\\  \langle\big[ c_{2},c_{1}\big] \rangle &  0 \end{pmatrix}
\end{align}
in terms of Majorana operators \eqref{maj} for this mode. Defining the $2 \times 2$ symplectic matrix
\begin{align*}
\Omega= \begin{pmatrix}0 & -1\\1& 0 \end{pmatrix}
\end{align*}
and  taking the product $\Omega \Gamma$, we find
\begin{align}\label{tr}
\operatorname{Tr}\Big( \Omega \Gamma \Big) = \ii  \langle\big[ c_{1},c_{2}\big] \rangle 
=\operatorname{Tr}( \ii[c_{1},c_{2}]\rho) 
\end{align}
The average energy $E(\rho) = \omega \operatorname{Tr}[\rho a^{\dagger}a]$, which  becomes
\begin{align}\label{nrg}
E(\rho) = \frac{ \omega}{2}  \operatorname{Tr}\Big[ \rho\big( c_{1}^{2} - \ii[c_{1},c_{2}] + c_{2}^{2}\big) \Big]
\end{align}
Substituting \eqref{tr} in \eqref{nrg} and taking note that $c_{1}^{2}= 1/2 = c_{2}^{2}$, we obtain $E(\Gamma) = \frac{\omega}{2} \Big(1-\operatorname{Tr}(\Omega \Gamma)\Big)$ as expected.  This is the average energy for a single mode of the state with frequency $\omega$.

As we consider non-interacting fermionic modes, the average energy of an $n$-mode state is defined as the sum of the average energy of each of the individual modes. In terms of covariance matrix this is given as
\begin{align}\label{nrgcov}
E(\Gamma_{n}) = \frac{\omega_{1}}{2}\Big(1-\operatorname{Tr}(\Omega_{1} \Gamma_{1})\Big) +\cdots+ \frac{\omega_{n}}{2}\Big(1-\operatorname{Tr}(\Omega_{n} \Gamma_{n})\Big) 
\end{align}
where the symplectic matrix for the entire system is $\Omega = \bigoplus_{j=1}^{n} \Omega_{j}$.

\section{Characterizing a Gaussian passive and non-passive fermionic state}
 
We are now ready to characterize quantum states with a covariance matrix $\Gamma$ for which the average energy \eqref{nrgcov} can be minimized by a Gaussian unitary transformation. 

\subsection{Standard form of a covariance matrix}

Let $\rho$ be the state of a two-mode system each with frequencies $\omega_{a}$ and $\omega_{b} \geq \omega_{a}$, and define $\Gamma$ as the covariance matrix of the two-mode system. 
Any two-mode covariance matrix can be brought to the form
\begin{align}\label{nf}
\Gamma_{sf} = \begin{pmatrix} 0 & a & 0& -e_{1} \\ -a & 0 & -e_{2} & 0\\ 0 & e_{2} & 0 & b\\ e_{1} & 0 & -b & 0 \end{pmatrix} 
\end{align}
 by a local orthogonal operation (LOO) $O_{\text{loc}} =  O_{\text{loc,a}} \oplus O_{\text{loc,b}}$, that is
$\Gamma_{sf} = O_{\text{loc}}\Gamma O^{T}_{\text{loc}}$ \cite{Jens2016}.

The more restrictive set of  pure Gaussian states are characterized by $\Gamma^{2}_{sf} = -\mathbb{1}$. This implies that the covariance matrix of the two-mode pure fermionic Gaussian state can be brought to the form
\begin{align}\label{nf1}
\Gamma_{sf} ^{p}= \begin{pmatrix} 0 & a & 0& -e \\ -a & 0 & -e& 0\\ 0 & e & 0 & a\\ e & 0 & -a& 0 \end{pmatrix} 
\end{align}
with $e=(1-a^{2})^{1/2}$ \cite{botero2004,Jens2016} so that the fermionic system depends only on one parameter $a$.

Now suppose we have a product of two fermionic modes with the covariance matrix in the standard form \eqref{nf}. Its average energy according to equation \eqref{nrgcov} is given as
\begin{align}\label{r1}
E(\Gamma_{sf}) & = \frac{\omega_{a} }{2}( 1 - 2a) + \frac{ \omega_{b}}{2}( 1 - 2b)
\end{align}
where $\omega_{a}$ and $\omega_{b}$ are the frequencies of the modes. We shall now
prove:

\begin{thm}\label{trm}
 Any (not necessarily Gaussian) state of two noninteracting fermionic modes with frequencies $\omega_{b} \geq  \omega_{a}$ is Gaussian-passive if and only if its covariance matrix $\Gamma$ is
\begin{enumerate}[(i)]
\item in Williamson standard form \cite{botero2004}
\begin{align}\label{sf}
\Gamma = \begin{pmatrix} 0 & a& 0& 0 \\ -a& 0 & 0 & 0\\ 0 & 0 & 0 & b \\ 0 & 0 & -b& 0 \end{pmatrix} 
\end{align} 
with $\lambda_{a} > \lambda_{b}$ for $\omega_{b} \neq  \omega_{a}$, or 
\item in the form
\begin{align}\label{st}
\Gamma = \begin{pmatrix} 0 & a& 0& -e \\ -a& 0 & e & 0\\ 0 & -e & 0 & b \\ e & 0 & -b& 0 \end{pmatrix} 
\end{align} 
for equal frequencies  $\omega_{b} = \omega_{a}$.
\end{enumerate}
\end{thm}

To prove this theorem, we start with the most general covariance matrix that any state $\rho$ may have and apply Gaussian operations to reduce its average energy until minimal. At this point we obtain a state $\rho'$ with minimal energy. We compare the energy of $\rho'$ with that of $\rho$ and identify under which conditions the energy of $\rho$ has been lowered. We thus can identify the characteristics of Gaussian-passive states from these conditions.  We consider here even fermionic systems
for which $\operatorname{Tr}(X) = 0$. As noted  in section \ref{FGx}, these have no Grassmann variables and so have vanishing first moment.

\subsection{Local Orthogonal Transformations}
We note that the covariance matrix $\Gamma$ of a two-mode fermionic system can be brought to its standard form through a local orthogonal transformation $O_{\text{loc}} = O_{\text{loc}_{a}} \oplus O_{\text{loc}_{b}}$, that is 
\begin{align}\label{loc}
O_{\text{loc}}\Gamma O_{\text{loc}}^{T}=\Gamma_{sf}= \begin{pmatrix}  A & E\\-E^{T}&B\end{pmatrix}
\end{align}
with 
\begin{align*}
O_{\text{loc}_{a}} =   \begin{pmatrix} \cos(\phi_{a}) & \sin(\phi_{a}) \\
-\sin(\phi_{a}) & \cos(\phi_{a})
\end{pmatrix}
\end{align*}
and where each element of  $\Gamma_{sf}$ is a $2 \times 2$ matrix 
\begin{align*}
A =  \begin{pmatrix}  0 & a\\-a&0\end{pmatrix},\quad  B =  \begin{pmatrix}  0 & b \\- b&0\end{pmatrix}, \quad E =  \begin{pmatrix}  0 & e_{1}\\ e_{2}& 0\end{pmatrix}.
\end{align*}
$A$ and $B$ describe the local covariance matrix of each mode and $E$ describes the correlation between the two modes. By inverting equation \eqref{loc}, we can write the local covariance matrix of a two-mode system as  
\begin{align}\label{trans1}
O^{T}_{\text{loc}}\Gamma_{sf} O_{\text{loc}}=\Gamma 
\end{align}
and we note that the inverse operations are also local orthogonal transformations. 

We ask here: given a state with a two-mode covariance matrix $\Gamma$,
%whose covariance matrix $\Gamma$ is of the form  \eqref{trans}, 
can work be extracted from the system? That is, can the average energy corresponding to $\Gamma$ be reduced?   To answer this question we compute the average energy $E(\Gamma)$ corresponding to the covariance matrix $\Gamma$ and find using \eqref{loc} that $E(\Gamma) = E(\Gamma_{sf})$ as given in equation \eqref{r1}. Since the energies are the same, it becomes clear that states with covariance matrix $\Gamma = \Gamma_{sf}$ are Gaussian passive under a local orthogonal transformation. 

However the energy of such states may be reduced by global orthogonal transformations, as we will show in the next section. 

\subsection{Two mode Squeezing}
Now suppose a state has a covariance matrix in the standard form \eqref{loc}. We have seen in the previous subsection   that such a state is Gaussian passive under a local orthogonal transformation. In this section, we will apply the  global orthogonal transformation \eqref{sq1} to   the system and see if its average energy can be reduced.  Computing the corresponding two-mode squeezed covariance matrix $\hat{\Gamma}_{TM}= S(r)\Gamma_{sf}S(r)^{T}$, we find

\begin{align}\label{nff}
\hat{\Gamma} _{TM}= \begin{pmatrix} 
0 & a' & 0& -e'_{1} \\ -a' & 0 & -e'_{2} & 0\\ 0 & e'_{2} & 0 & b'\\ e'_{1} & 0 & -b' & 0 
\end{pmatrix} 
\end{align}
where 
\begin{subequations}\label{mat}
\begin{align}
a'&=a c_{r}^{2} - b s^{2}_{r} - \frac{1}{2}(e_{1}+e_{2})s_{2r},\\
b'&=-a s^{2}_{r} +bc^{2}_{r}-\frac{1}{2}(e_{1}+e_{2})s_{2r} \\
 e'_1&=\frac{1}{2}(a+b)s_{2r}+e_{1}c^{2}_{r} - e_{2}s^{2}_{r}\\
  e'_2&=\frac{1}{2}(a+b)s_{2r}+e_{2}c^{2}_{r} - e_{1}s^{2}_{r}
\end{align}
\end{subequations}
with $c_{r} = \cos(r)$ and $s_{r}=\sin(r)$ respectively. To see if this transformation can reduce the average energy, we compute $E(\hat{\Gamma}_{TM})$ using \eqref{endef}, obtaining
\begin{align}\label{enrgy}
E(\hat{\Gamma}_{TM}) = \frac{\omega_{a}}{2}(1-2a') + \frac{\omega_{b}}{2}(1-2b')
\end{align}
and substituting equations \eqref{mat} into \eqref{enrgy}, we get
\begin{align}\label{etm1}\nonumber
E(\hat{\Gamma}_{TM}) &=\omega_{a}\Big[b\sin^{2}(r) - a\cos^{2}(r)\Big]+\omega_{b} \Big[a\sin^{2}(r) - b\cos^{2}(r)\Big]\\
&+\frac{( \omega_{a} + \omega_{b}) }{2} \Big[1+(e_{1}+e_{2})\sin(2r)\Big]
\end{align}
and minimizing this with respect to the squeezing parameter $r$ we find  the condition
\begin{align}\label{condition}\nonumber
\frac{\partial }{\partial r }E(\hat{\Gamma}_{TM})=& 0\\
 \Rightarrow (a+b)\sin(2r) + (e_{1}+e_{2})\cos(2r) =& 0
\end{align}
whose solution is
\begin{align}\label{rr}
r =-\frac{1}{2} \tan^{-1}\Bigg[\frac{e_{1}+e_{2}}{(a+b)}\Bigg] =-\frac{1}{2} \tan^{-1}(\lambda)
\end{align}
where $\lambda = (e_1+e_2)/(a+b)$. The minimized energy is
\begin{align} 
E_{\text{min}}(\hat{\Gamma}_{TM}) &= \frac{( \omega_{a} + \omega_{b}) }{2}\Big(1-(a+b)\sqrt{1+\lambda^2} \Big)  \nonumber\\
&\quad +\frac{1}{2}( \omega_{b} - \omega_{a})(a-b).
\label{etm} 
\end{align}
Define $e = (e_{1}-e_{2})/2$, the elements of the covariance matrix \eqref{mat} are now  
\begin{subequations}
\begin{align}\label{ae}
\tilde{a}'&=\frac{(a+b)}{2}\sqrt{1+\lambda^2} + \frac{(a-b)}{2}\\\label{be}
\tilde{b}'&=\frac{(a+b)}{2}\sqrt{1+\lambda^2} -\frac{(a- b)}{2} \\\label{e1}
\tilde{ e}'_1&= e, \quad  \tilde{e}'_2 = -e
\end{align}
\end{subequations}

 We pause to comment on the interpretation of these matrix elements.  In addition to minimizing the system's average energy, the squeezing parameter \eqref{rr} reduces the off-diagonal elements in \eqref{nff} to a single parameter $e$ so that the resulting covariance matrix is of the form
 \begin{align}\label{nfh}
\Gamma_{GP} = \begin{pmatrix} 
0 & \tilde{a}' & 0& -e\\ -\tilde{a}' & 0&e & 0\\ 0 & -e & 0 &\tilde{ b}'\\ e& 0 & -\tilde{b}'& 0 
\end{pmatrix}
\end{align}

 If the state is a two-mode pure fermionic Gaussian state whose covariance matrix is of the form \eqref{nf1}, the  two-mode squeezing operation takes the state's covariance matrix to the form
 \begin{align}\label{pfgs}
\Gamma_{GP}^{p} = \begin{pmatrix} 
0 & 1 & 0& 0\\ -1 & 0&0 & 0\\ 0 & 0& 0 & 1\\ 0 & 0 & - 1& 0 
\end{pmatrix}
\end{align}
with property $(\Gamma_{GP}^{p})^{2} = -\mathbb{1}$. This corresponds to the covariance matrix of a pure fermionic Gaussian state in the Williamson normal form \cite{botero2004}. To achieve a Williamson normal form covariance matrix for the general two-mode fermionic mode, we consider further Gaussian unitary transformations on the system.
 
 \subsection{Beam Splitting}
 The last Gaussian operation we have to consider is the beam splitting operation. This transformation on fermionic phase space is represented by the transformation matrix \eqref{bs1}.
We find 
\begin{align}
\Gamma_{BS}= B(\theta) \hat{\Gamma}_{GP} B^{\dagger}(\theta) =
\begin{pmatrix} 0 & A&0&D\\ - A& 0& -D& 0\\ 0& D&0&B\\ -D&0&-B&0\end{pmatrix}
\end{align}
where
\begin{subequations}\label{mat1}
\begin{align}
A&= \tilde{a}' \cos^{2}{\theta} + \tilde{b'}\sin^{2}(\theta)+e \sin(2 \theta)\\
B&= \tilde{b}' \cos^{2}{\theta} + \tilde{a'}\sin^{2}(\theta)-e\sin(2 \theta)\\
 D&=\frac{1}{2}(\tilde{a}'-\tilde{b}')\sin{2\theta} -e\cos(2\theta)
\end{align}
\end{subequations}
The average energy corresponding to $\Gamma_{BS}$ is
\begin{align}\label{bsm}\nonumber
E(\Gamma_{BS}) = & -\omega_{a}\Big[\tilde{b}' \sin^{2}(\theta) +\tilde{a}' \cos^{2}(\theta)\Big]
\nonumber\\
&\quad -\omega_{b} \Big[\tilde{a}' \sin^{2}(\theta) + \tilde{b}' \cos^{2}(\theta)\Big] \\
&+\frac{( \omega_{a} + \omega_{b}) }{2} + ( \omega_{b} - \omega_{a}) e\sin(2\theta)
\end{align}
Again energy is minimized for the value of $\theta$ satisfying the equation
\begin{align} 
(\omega_{b}-\omega_{a})\Bigg[(\tilde{b}' - \tilde{a}')\sin(2 \theta) +2e\cos(2\theta)\Bigg]= 0
\end{align}
implying
\begin{align*}
\theta = -\frac{1}{2}\tan^{-1}\Big(\frac{2e}{\tilde{b}'-\tilde{a}'} \Big)= -\frac{1}{2}\tan^{-1}\mu
\end{align*}
where $\mu = 2e/(\tilde{b}' - \tilde{a}')$. The minimized energy under the beam splitting operation is then
\begin{align} 
E_{\text{min}}(\hat{\Gamma}_{BS}) &= \frac{( \omega_{b} - \omega_{a}) }{2}\Big((\tilde{a}'-\tilde{b}')\sqrt{1+\mu^2} \Big)  \nonumber\\
&\quad +\frac{1}{2}( \omega_{b} + \omega_{a})\Big(1-(\tilde{a}'+\tilde{b}')\Big),
\label{beams} 
\end{align}
 and the corresponding minimized matrix element is 
\begin{subequations}
\begin{align}\label{ae1}
A&=\frac{(\tilde{a}'+\tilde{b}')}{2} + \frac{(\tilde{a}'-\tilde{b}')}{2}\sqrt{1+\mu^2} \\\label{be1}
B&=\frac{(\tilde{a}'+\tilde{b}')}{2} -\frac{(\tilde{a}'-\tilde{b}')}{2}\sqrt{1+\mu^2} \\\label{e2}
D&=0
\end{align}
\end{subequations}
For equal frequencies $\omega_{a} = \omega_{b}$, the average energy is unchanged, that is $E_{\text{min}}(
\hat{\Gamma}_{TM}) = E_{\text{min}}(\hat{\Gamma}_{BS})$ and we conclude that the state with covariance matrix \eqref{nfh} is Gaussian passive.  However for different frequencies assume w.l.o.g. that $\omega_{b}>\omega_{a}$, the covariance matrix for the minimized state under beam splitting operation is in the Williamson normal form \cite{botero2004}
\begin{align}\label{beams1}
\Gamma_{GP}^{1}= \begin{pmatrix} 0 & A &0&0\\ - A& 0& 0& 0\\ 0& 0&0&B\\ 0&0&-B&0\end{pmatrix},
\end{align}
 with eigenvalues given as $\lambda_{a}= \pm \ii A$ and $\lambda_{b} = \pm \ii B$. If $a>b$, we find that $\lambda_{a}>\lambda_{b}$ and so the lower frequency mode has the higher population.

 We see that the effect of the orthogonal transformation on the fermionic two-mode covariance matrix is to decompose the modes and bring them into a product of single-mode locally thermal states diagonal in the Fock basis. An example of Gaussian passive state of two modes with different frequencies is that of a product of single mode thermal states, in which each mode has different temperature. In this case, the Williamson eigenvalues are $\lambda_{i} = \tanh \Big( \frac{\omega_{i}}{2T_{i}}\Big)$. For $T_{b} \neq 0$ the condition $\lambda_{a} >\lambda_{b}$ for Gaussian passivity can be expressed as
  \begin{align}
  \frac{\omega_{a}}{\omega_{b}} > \frac{T_{a}}{T_{b}}
  \end{align}

  As shown in section \ref{act}, within the framework of  general operations, the product of two thermal states at different temperature is passive, regardless of the frequencies of the modes involved.  And from above, we see that such a state is also Gaussian passive showing that all passive states are obviously Gaussian passive, but the converse may not be true \cite{eric2016} as we will show in the next section.
  
\subsection{More General Operations}

So far we have focused on characterizing a general fermionic state according to whether work can be extracted or not using Gaussian unitary transformations. We started with the covariance matrix of a general two-mode fermionic system, applied gaussian unitary operations to extract the energy from the system and then we arrived at the Gaussian passive state \eqref{beams1}, where no further energy could be extracted by additional Gaussian unitary transformation. A reasonable question then arises:  in the process of characterizing a (not necessarily Gaussian) state, how much extractable work is sacrificed by using  Gaussian unitary transformations  instead of general unitary transformations? To address this question we will follow a procedure similar to that  in the bosonic case \cite{eric2016}.

 In the  characterization  process  we fixed the second moment of the fermionic state, which only uniquely identifies a state if it is Gaussian. Two steps   therefore lead us to answering the above question.  1) First we must find a non-Gaussian state that is compatible with a given Gaussian passive state, or in other words we must find  a non-Gaussian state with the same second moment as that of the Gaussian passive state. 2) We must show that a general unitary transformation on the resulting non-Gaussian state can  lower its energy to the minimal value.

 To proceed, we first note that the covariance matrix of a general two-mode Gaussian-passive state \eqref{beams1} is  identical to the covariance matrix of a product of locally thermal states of  two different modes each with different effective temperatures. One could then consider a single fermionic mode in a thermal state with arbitrary temperature and then  find a pure state whose  second moment is that of this single mode thermal state. Then one could certainly find pairs of states of this kind whose tensor product is compatible with a Gaussian-passive locally thermal two-mode state.
For example, in the Fock basis, the fermionic state
\begin{align}\label{psi-notexist}
\ket{\psi} = \sqrt{1-p}\ket{0} +\sqrt{ p}\ket{1}, \qquad 0\leq p\leq 1
\end{align}
has a covariance matrix of the form
\begin{align}\label{srule}
\begin{pmatrix}
0& 2p-1\\
1-2p&0
\end{pmatrix}
\end{align}
and so  by carefully choosing the continuous parameter $p$, we can bring the covariance matrix to look like that of a single-mode fermionic thermal state  with inverse temperature $\beta$
\begin{align}
\Gamma_{th} = \begin{pmatrix} 0 & \tanh\Big( \frac{\beta \omega}{2}\Big) \\-\tanh\Big( \frac{\beta \omega}{2}\Big) & 0 ,\end{pmatrix}
\end{align}
where $\omega$ is the mode frequency.
Unfortunately, the state \eqref{psi-notexist}  is prohibited by a super-selection rule \cite{Aharonov1967} and so does not exist. 

However, another example would be the fermionic vacuum state $\ket{0}$ and a single fermion state $\ket{1}$ each having covariance matrices
\begin{align*}
\Gamma_{\rho_{0}}= \begin{pmatrix}
0& 1\\
-1 & 0
\end{pmatrix}, \quad \Gamma_{\rho_{1}}= \begin{pmatrix}
0& -1\\
1 & 0
\end{pmatrix}, 
\end{align*}
respectively.  Given that the $(\ket{1}, \ket{0})$ states are pure, their covariance matrices satisfy the condition $\Gamma_{\ket{i}}^{2} = -\mathbb{1}$. We define the free energy of these states as
\begin{align}
F(\rho) =E(\rho) - TS(\rho),
\end{align}
where $S(\rho) = - \text{Tr}[\rho \text{ln} (\rho)]$ is the von Neumann entropy which is vanishing for pure states, and $E(\rho) $ is the average (internal) energy.

 Now to achieve our first task, consider pairs of the single fermionic systems encoded into a bipartite Hilbert space $\mathcal{H}_{ab} = \mathcal{H}_{a}\otimes \mathcal{H}_{b}$ of subsystems $a$ and $b$ respectively. The state is defined by a density operator $\rho_{ab}^{1} = \ket{00}_{ab}\langle 00|$ and $\rho_{ab}^{2}=\ket{11}_{ab}\langle 11| $ respectively, the resulting states correspond to direct sum of locally pure fermionic Gaussian states. Their covariance matrices are respectively 
\begin{align*}
\Gamma_{\rho_{ab}^{1}}= \Gamma_{\rho_{1}}^{a}\oplus \Gamma_{\rho_{1}}^{b}, \quad \Gamma_{\rho_{ab}^{2} }= \Gamma_{\rho_{2}}^{a}\oplus \Gamma_{\rho_{2}}^{b},
\end{align*}
which is the same as the CM of pure fermionic Gaussian passive state \eqref{pfgs}. For our second task, given that the constructed states are pure, their free energy is thus identical to the average energy. Interestingly there is no way to lower the average energy of the constructed state $\rho_{ab}^{1}$, however the energy of the state $\rho_{ab}^{2}$ can be lowered by applying a (non Gaussian) unitary transformation that takes the pure state to vacuum state.  This shows that $\rho_{ab}^{2}$ is Gaussian passive but not passive while $\rho_{ab}^{1}$ is both passive and Gaussian passive, as expected.

\section{Conclusion}

We have investigated the problem of work extraction from fermionic systems, finding a number of similarities and differences with their bosonic counterparts.
 
Thermal states at positive temperatures are the only completely passive states from which work cannot be extracted no matter how many available copies \cite{carlo2017,CPV2015,Pusz1978}. 

Any quantum state out-of-equilibrium is a potential resource for work extraction. However for fermions the situation is somewhat subtle. We have shown that  under arbitrary unitary transformations there is no way to process a product of two fermionic modes in a thermal state to extract work, independent of mode temperatures and frequency. This is quite unlike the situation the bosonic counterpart \cite{eric2016}, and
suggests that fermionic systems are not as useful for quantum thermodynamic applications such as construction of quantum heat engines \cite{Tien2004}. However we found that a product of more than two fermionic modes in different thermal states  was non-passive (under a certain temperature constraint), implying
work extraction is possible in this system. The challenge of  generating the necessary unitary operation for this work extraction could be a limitation.

We extended the notion of Gaussian passivity to fermionic systems and presented criteria for identifying  fermionic states according to their Gaussian (non-Gaussian) passivity; that is, according to our ability (inability) to extract work from them using Gaussian unitary transformations. This characterization is based on the second statistical moment of the two-mode fermionic system, which is known to have complete information about the system.  This implies that our characterization provides information about the Gaussian ergotropy of the system (that is the maximum extractable energy in a Gaussian unitary process). Our result showed that under non-Gaussian (general) unitaries, we showed that work can be extracted from a general two-mode fermionic system.

There is still much that can be done with Fermionic Gaussian systems.  A classification of their dynamics for open systems (analogous to the bosonic case \cite{Dan2018}) remains to be carried out, along with their time evolution under rapid bombardment.  Work on these topics is in progress.

%\tcb{\bf [How much of this next  paragraph do we really need?]} 
% 
%\tcr{To summarize, we can say that work can be generally obtained from passive states if any unitary operation can be performed on the system. On the other hand, work can be performed on thermal states if unitary operations generated from quadratic Hamiltnoian is performed on the system. This shows that Gaussian transformations which corresponds to unitary generated from quadratic Hamiltonian is in general more constraint than the set of general unitary operations.}
%
%\tcr{Many limitations of Gaussian operations are known both in quantum computation \cite{} and quantum thermodynamics \cite{} where } \tcb{\bf [Need to finish this sentence!]}
%

\section*{Acknowledgements}
This work was supported in part by the Natural Science and Engineering Research Council of Canada.  We are grateful to Eric Brown for helpful discussion and correspondence.

\bibliography{references}
\end{document}